\documentclass[twocolumn,aps,floatfix,showpacs,prb,superscriptaddress]{revtex4}
\usepackage{epsfig}
\usepackage{graphicx}
\usepackage{amsmath}
\usepackage{amssymb}
\usepackage{amsfonts}
\usepackage{tabularx}
\usepackage{color}
\usepackage{dcolumn}
\usepackage{bm}
\usepackage{floatflt}
\usepackage{setspace}
\usepackage{subfigure}

\begin{document}

\title{Sum-rules for electron energy-loss near-edge spectra}

\author{J\'{a}n Rusz} \email{jan.rusz@fysik.uu.se}
\affiliation{Department of Physics, Uppsala University, Box 530, S-751 21 Uppsala, Sweden}
\affiliation{Institute of Physics, Academy of Sciences of the Czech Republic, Na Slovance 2, CZ-182 21 Prague, Czech Republic}
\author{Olle Eriksson}
\affiliation{Department of Physics, Uppsala University, Box 530, S-751 21 Uppsala, Sweden}
\author{Pavel Nov\'{a}k}
\affiliation{Institute of Physics, Academy of Sciences of the Czech Republic, Na Slovance 2, CZ-182 21 Prague, Czech Republic}
\author{Peter M. Oppeneer}
\affiliation{Department of Physics, Uppsala University, Box 530, S-751 21 Uppsala, Sweden}

\date{\today}
\begin{abstract} 
We derive four sum-rule expressions for spectra measured in electron energy-loss near edge structure experiments. These sum-rules permit the determination spin and orbital magnetic moments, spin-orbit interaction and number of states, analogously to the sum rules of x-ray magnetic circular dichroism.
The derivation of the sum-rules is based on dynamical electron diffraction theory and the properties of the mixed dynamic form-factor. The accuracy of the sum-rules is tested by a complete evaluation of the thickness dependent electron energy-loss spectra for iron, cobalt, and nickel crystals. We find that the sum-rules reproduce both spin and orbital moments with very good accuracy. Our results provide a 
foundation for the use of the energy loss magnetic chiral dichroism technique as a quantitative probe of element specific magnetic properties.
\end{abstract}


\maketitle

X-ray magnetic circular dichroism\cite{laan,schutz,chen} (XMCD) is a powerful and widely used spectroscopic technique for the investigation of magnetic materials. The strength of the XMCD stems particularly from the well-known sum-rules, which enable one to obtain element specific information on spin and orbital magnetic moments in the material from an integration of the measured XMCD spectrum. \cite{thole,carra,ankudinov}

The existence of the electron analogue to the X-ray radiation based XMCD was recently demonstrated by Schattschneider {\it et al}.\cite{nature} An electron beam passing through a magnetic material in an electron transmission microscope (TEM) was proved to exhibit a magnetic dichroism phenomenon, which has been named energy-loss magnetic chiral dichroism (EMCD). The EMCD technique has the advantage that, due to the short de Broglie wavelength of the electrons, an excellent lateral resolution can be achieved. Modern TEMs easily reach sub-nanometric lateral resolutions, though this has not yet been achieved for EMCD. By improvements of the experimental geometries \cite{lacbed,lacdif} the signal-to-noise ratio (SNR) of the EMCD has recently been significantly enhanced. Lateral resolutions of 10 nm and below have been reached with a very good SNR, exceeding thereby the resolution accomplished in current XMCD experiments.

In spite of the encouraging recent achievements of the EMCD, there is a lack of quantitative interpretation of the measured spectra connecting these to magnetic properties of the studied material. Here we fill this gap by deriving spin and orbital momentum sum-rule expressions for the EMCD technique. Our results thus provide a formalism to establish this new experimental method as a {\it quantitative probe} which provides {\it element specific information on spin and orbital magnetism.}

EMCD is essentially an electron energy-loss near-edge structure (ELNES) experiment, in which a core state electron is excited to an unpopulated valence state above the Fermi level. Precise evaluation of the double differential scattering cross-section [DDSCS, $\sigma(E)$], which is an output of ELNES experiment, requires the combination of dynamical diffraction theory and microscopic calculations.\cite{papertheory}

In the following we present the derivation of the EMCD sum-rules. It was demonstrated that the dichroic effect is related to the imaginary part of the mixed dynamic form factor [MDFF; usually denoted as $S(\mathbf{q},\mathbf{q'},E)$ -- a function of energy and momentum transfer vectors].\cite{nature,lacbed} Because of its hermiticity, the imaginary part of the MDFFs can be separated out by performing two measurements -- similarly to XMCD -- in which the order of $\mathbf{q},\mathbf{q'}$ is reversed. It is assumed that both measurements are performed in symmetrical geometry, so that the dynamical diffraction equations lead to the same set of Bloch coefficients and Bloch-wave vectors. For example, a reversal of the sample magnetization without changing the experimental geometry would provide such two measurements.\cite{sysrow} Then the difference of two such spectra, let's denote them by $\sigma_+(E)$ and $\sigma_-(E)$, is determined solely by the imaginary parts of contributing MDFFs. The sum of these spectra is a function of the real parts of MDFFs.

The MDFF itself is given by the following expression
\begin{equation}
 S(\mathbf{q},\mathbf{q'},E) = \sum_{i,f} \langle i|e^{-i\mathbf{q}\cdot\mathbf{r}}|f\rangle \langle f|e^{i\mathbf{q'}\cdot\mathbf{r}}|i\rangle \delta(E-E_{if})
\end{equation}
where the sum is taken over initial (core) states and final (unoccupied) band states. The sum-rules are expressed in terms of integrals of the signal from a particular edge over the whole energy spectrum. Owing to the $\delta$-function the integration of the MDFF over energy is trivial and the sum over final states can be formally expressed as
\begin{equation}
  \sum_f^{unocc} | f \rangle \langle f| = \hat{\openone} - \sum_f^{occ} | f \rangle \langle f| = \hat{\openone} - \hat{\rho}
\end{equation}
where $\hat{\rho}$ is the ground state density matrix.

The next step is to consider only dipole allowed allowed transitions $l \to l + 1 \equiv L$ (e.g.\ $2p \to 3d$ in transition metals). All such transitions are contained in the second term of the Rayleigh expansion of the exponential $e^{i\mathbf{q}\cdot\mathbf{r}}$, which is equal to $3j_1(qr)\mathbf{q}\cdot\mathbf{r}/(qr)$ [$j_\lambda(x)$ is a spherical Bessel function of order $\lambda$]. For small $qr$ this expression reduces to a simple scalar product $\mathbf{q}\cdot\mathbf{r}$ appearing also in the Taylor expansion of the exponential (and often used in XMCD interpretation). But for $qr \gtrsim 1$, which is a common situation in ELNES, these expressions are very different\cite{papertheory} and the Rayleigh expansion provides significantly higher accuracy. The second assumption is that the radial part of the core state wavefunction is spin and $j$-independent. We note that an exact sum rule for the orbital momentum in the XMCD has been derived,\cite{kunes} and in Refs.~\onlinecite{ankudinov,spinsr} it was demonstrated how to include spin and $j$-dependence. In principle these approaches can be extended to the EMCD sum-rules derived below, but for clarity of presentation we will assume that these approximations are accurate enough.

Under these assumptions the integrated MDFF (we use the same notation for energy-integrated MDFF omitting the energy parameter) becomes
\begin{eqnarray} \label{eq:mdffap}
\lefteqn{S(\mathbf{q},\mathbf{q'}) \approx 12\pi\frac{l+1}{2l+1}\langle j_{1}(q)\rangle_{lL}\langle j_{1}(q')\rangle_{lL}} \nonumber \\
 & \times & \sum_{\mu,\mu'}Y_{1}^{\mu}(\hat{q})^{\star}Y_{1}^{\mu'}(\hat{q}') \nonumber \\
 & \times & \sum_{mm'ss'}\begin{pmatrix}l & 1 & L \\
-m & \mu & m-\mu\end{pmatrix}\begin{pmatrix}l & 1 & L \\
-m' & \mu' & m'-\mu'\end{pmatrix} \nonumber \\
 & \times & [l+\delta_{J,+} + 2(\delta_{J+}-\delta_{J-})\langle l m s | \hat{\mathbf{s}}\cdot\hat{\mathbf{l}} | l m' s' \rangle] \nonumber \\
 & \times & \langle L,m-\mu,s|\hat{\openone}-\hat{\rho}_L|L,m'-\mu',s' \rangle
\end{eqnarray}
where $l$ and $L$ are the orbital quantum numbers of initial and final states, respectively, $\hat{\rho}_L$ is the $L$-projected part of the density matrix, $\langle j_{1}(q)\rangle_{lL}$ is the integral of the spherical Bessel function, $j_1(qr)$, and radial parts of initial and final wavefunctions, and $Y_l^m(\hat{q})$ are the spherical harmonics, $\hat{q}=\mathbf{q}/q$. The Wigner $3j$-symbols appear due to integrals of spherical harmonics in the real space (for details see Ref.~\onlinecite{papertheory}). Kronecker deltas $\delta_{J,\pm}$ select the particular edge ($J = l \pm \frac{1}{2}$). This expression is an ELNES analogy of Eq.~(5) in Ref.~\onlinecite{ankudinov}. The formal similarities of this expression and matrix elements for the x-ray absorption spectra allow to use the same $\hat{L}$-operator algebra as in Ref.~\onlinecite{ankudinov}. Note, however, that in contrast to XMCD, where the particular photon polarization $\mu$ can be chosen in a particular experiment, in the ELNES expression every MDFF term is a combination of all three ``polarizations'' including cross-terms between them (non-diagonal terms in $\mu \ne \mu'$).

The orbital sum-rule is given as a sum of the integrated dichroic signal over both $J = l \pm \frac{1}{2}$ edges. In such case the last-but-one line of Eq.~(\ref{eq:mdffap}) (the only $J$-dependent part) reduces to $(2L-1)$. After evaluation of all matrix elements and summation over $\mu,\mu'$ we obtain
\begin{equation} \label{eq:orbsr}
  \sum_J \textrm{Im}[S(\mathbf{q},\mathbf{q'})] = \frac{1}{2}M_L(q,q')(2L-1) \mathbf{Q} \cdot \langle \hat{\mathbf{L}} \rangle
\end{equation}
where
\begin{equation}\label{eq:mlqq}
M_L(q,q')=\frac{9}{(2L-1)(2L+1)} \frac{\langle j_{1}(q)\rangle_{lL}\langle j_{1}(q')\rangle_{lL}}{q q'}
\end{equation}
and $\mathbf{Q}=\mathbf{q}\times\mathbf{q'}$.

An EMCD analogy of the XMCD spin sum-rule is
\begin{eqnarray} \label{eq:spinsr}
  \lefteqn{\frac{1}{L}\left. \textrm{Im}[S(\mathbf{q},\mathbf{q'})]\right|_{J+} -
  \frac{1}{L-1}\left. \textrm{Im}[S(\mathbf{q},\mathbf{q'})]\right|_{J-} =} \nonumber \\
 & = & \frac{1}{3} M_L(q,q') (2L-1) \mathbf{Q} \cdot  \left[ \langle \hat{\mathbf{S}} \rangle + \frac{2L+3}{L} \langle \hat{\mathbf{T}} \rangle \right]
\end{eqnarray}
where $\hat{\mathbf{T}}$ is the the magnetic dipole operator. 

Equations~(\ref{eq:orbsr}) and (\ref{eq:spinsr}) fully disclose the information contained in the energy integral of the imaginary part of MDFF, which can be re-formulated as
\begin{eqnarray}
  \lefteqn{\left. \textrm{Im}[S(\mathbf{q},\mathbf{q'})]\right|_{J\pm}  =  M_L(q,q')} \nonumber \\
 & \times & \mathbf{Q} \cdot  \left[ \frac{L-\delta_{J-}}{2} \langle \hat{\mathbf{L}} \rangle \pm \frac{L(L-1)}{3} \left[ \langle \hat{\mathbf{S}} \rangle + \frac{2L+3}{L} \langle \hat{\mathbf{T}} \rangle \right] \right] \nonumber \\
\end{eqnarray}
depending on particular edge $J=l \pm \frac{1}{2}$.

Because the real part of the MDFF does not change upon time inversion (in the dipole approximation, otherwise it only holds if the system also has inversion symmetry), it must be an even function of magnetic moments. It is possible to derive two other sum-rules for the real part of the MDFF. The first one is an analogy to the so called $N$ sum-rule,\cite{ankudinov} corresponding to the number of states
\begin{eqnarray}\label{eq:nsr}
  \lefteqn{\sum_J \textrm{Re}[\tilde{S}(\mathbf{q},\mathbf{q'})] = M_L(q,q') L^2} \nonumber \\
  & \times & \left[ (\mathbf{q}\cdot\mathbf{q'}) \frac{2(2L+1)(2L-1)}{3L} - (\mathbf{q}\cdot\mathbf{q'}) \langle \hat{\openone} \rangle \right. \nonumber \\
  & + & \left. \frac{1}{2L^2} \langle (\mathbf{q}\cdot\hat{\mathbf{L}})(\mathbf{q'}\cdot\hat{\mathbf{L}}) + 
                                      (\mathbf{q'}\cdot\hat{\mathbf{L}})(\mathbf{q}\cdot\hat{\mathbf{L}}) \rangle \right].
\end{eqnarray}
This is the only sum-rule, where a nonzero contribution arises from the unity operator in Eq.~(\ref{eq:mdffap}) -- leading to the first term of Eq.\ (\ref{eq:nsr}). The second term is proportional to the trace of the density matrix, i.e.\ the number of $L$-electrons, $N_e = \langle \hat\openone \rangle = \mathrm{Tr} [\hat\rho_L]$.

The second sum-rule for the real part of the MDFF is an analogy of the spin-orbit sum-rule\cite{ankudinov}
\begin{eqnarray} \label{eq:sosr}
  \lefteqn{\frac{1}{L}\left. \textrm{Re}[S(\mathbf{q},\mathbf{q'})]\right|_{J+} -
  \frac{1}{L-1}\left. \textrm{Re}[S(\mathbf{q},\mathbf{q'})]\right|_{J-} =} \nonumber \\
 & - & 2 M_L(q,q') \left[ (\mathbf{q}\cdot\mathbf{q'}) \langle \hat{\mathbf{L}}\cdot\hat{\mathbf{S}} \rangle \right.+ \nonumber \\
 & + & \frac{\langle (\mathbf{q}\cdot\hat{\mathbf{L}})(\mathbf{q'}\cdot\hat{\mathbf{S}}) + 
                                (\mathbf{q'}\cdot\hat{\mathbf{L}})(\mathbf{q}\cdot\hat{\mathbf{S}}) \rangle}{2(L-1)}  - \nonumber \\
 & - & \frac{\langle
   (\mathbf{q}\cdot\hat{\mathbf{L}})(\hat{\mathbf{L}}\cdot\hat{\mathbf{S}})(\mathbf{q'}\cdot\hat{\mathbf{L}}) + 
   (\mathbf{q'}\cdot\hat{\mathbf{L}})(\hat{\mathbf{L}}\cdot\hat{\mathbf{S}})(\mathbf{q}\cdot\hat{\mathbf{L}}) 
                         \rangle}{2L(L-1)} \Big ]
\end{eqnarray}
where the dominant term is proportional to the mean value of the spin-orbit operator.

In both these sum-rules there are some extra terms, which do not appear in their XMCD analogues. It is possible to show that these vanish in the limit of a strong enough magnetic field, where all magnetic moments are aligned along the $z$-axis (magnetization axis) and small energy losses, when the $z$-component of momentum transfer vectors vanishes.

These four sum-rules, written in a \textit{rotationally invariant} form, are a complete set of ELNES sum-rules within the dipole approximation. Ankudinov and Rehr\cite{ankudinov} derived six X-ray absorption spectroscopy (XAS) sum-rules, because there it is possible to prepare incoming light in a particular polarization ($+$, $-$ or $0$). In the ELNES experiment every measurement is in general a combination of all possible polarizations including `cross-terms'. Having two symmetric measurements, it is possible to set apart the signals which originate from the real or imaginary parts of the MDFF's. By separating the information obtained for the two edges we have four independent integrals of signals, which allow to construct the above-given four sum-rules. Expressed in simplifying terms, one can say that (i) the ELNES $N$ sum-rule is a combination of XMCD $N$ and anisotropic orbit sum-rules and (ii) the ELNES spin-orbit sum-rule is a combination of XMCD spin-orbit and anisotropic spin magnetic\cite{ankudinov} sum-rules, respectively.

To calculate the total DDSCS, we need to evaluate a sum over many MDFFs.\cite{papertheory} It is important to note that independent of the values of $\mathbf{q},\mathbf{q'}$, the MDFF is always given as a mean value of particular operators appearing in sum-rules. The summation then introduces only a prefactor (although a rather complex one). The complete orbital sum-rule is
\begin{eqnarray}
 \sum_{J}\frac{d\Delta\sigma}{d\Omega} & = & \frac{4\gamma^2}{a_0^2} \frac{\chi_f}{\chi_i} \sum_{\genfrac{}{}{0pt}{}{jlj'l'}{\mathbf{ghg}'\mathbf{h}'}}
        T_{jlj'l'} Y_{\mathbf{ghg}'\mathbf{h}'}^{jlj'l'} \sum_\mathbf{u} e^{i(\mathbf{q}-\mathbf{q}').\mathbf{u}} \nonumber \\
  & \times & \frac{M_L(q,q')}{2q^2q'^2}  (2L-1) (\mathbf{q}\times\mathbf{q'}) \cdot \langle \hat{\mathbf{L}} \rangle
\end{eqnarray}
and spin sum-rule
\begin{eqnarray}
 \lefteqn{\frac{1}{L}\left.\frac{d\Delta\sigma}{d\Omega}\right|_{J+} - \frac{1}{L-1}\left.\frac{d\Delta\sigma}{d\Omega}\right|_{J-} = } \nonumber \\
  & = & \frac{4\gamma^2}{a_0^2} \frac{\chi_f}{\chi_i} \sum_{\genfrac{}{}{0pt}{}{jlj'l'}{\mathbf{ghg}'\mathbf{h}'}}
        T_{jlj'l'} Y_{\mathbf{ghg}'\mathbf{h}'}^{jlj'l'} \sum_\mathbf{u} e^{i(\mathbf{q}-\mathbf{q}').\mathbf{u}} \nonumber \\
  & \times & \frac{M_L(q,q')}{3q^2q'^2} (2L-1) (\mathbf{q}\times\mathbf{q'}) \cdot  \left[ \langle \hat{\mathbf{S}} \rangle + \frac{2L+3}{L} \langle \hat{\mathbf{T}} \rangle \right] \quad
\end{eqnarray}
where $\Delta\sigma = \int [\sigma_+(E)-\sigma_-(E)]dE$ is an integrated difference signal from a particular edge.
All quantities entering the prefactor are accessible either from dynamical diffraction theory [$Y_{\mathbf{ghg}'\mathbf{h}'}^{jlj'l'}$ is a product of Bloch coefficients, $T_{jlj'l'}$ depends on sample thickness and Bloch wave-vectors] or from simple electronic structure calculations [factor $M_L(q,q')$, Eq.~(\ref{eq:mlqq})] ; they can be determined with good accuracy. More details can be found in our recent publication, Ref.~\onlinecite{papertheory}.

If we assume that $Q_{x,y} \ll Q_z$, which means that we neglect the $z$ components of $\mathbf{q},\mathbf{q'}$ (physically this means, that the electron energy-loss is negligible when compared to acceleration voltage), only $Q_z$ times the mean value of the $z$-component of operators will remain. However, for example, in $2p \to 3d$ transitions of iron, cobalt or nickel at 200~kV acceleration voltage the $z$-component of $\mathbf{q}$-vectors is approximately $1/3$ of the reciprocal lattice vectors, i.e.\ non-negligible. On the other hand, in systems with weak spin-orbit interaction or in strong enough external magnetic field, the magnetic moments in the sample will have a direction dictated by the external magnetic field, which is usually the optical axis of the transmission electron microscope (the $z$ axis). Then again one can approximate $\mathbf{Q} \cdot \langle \hat{\mathbf{L}} \rangle \approx Q_z \langle \hat{L}_z \rangle$.

If none of these assumptions hold, the sum-rules give a \emph{projection} of the spin, orbital and magnetic dipole moments on the direction of $\mathbf{Q}$. The following reasoning then helps to interpret the sum-rules predictions.
The most common EMCD geometries can be very well simulated using the systematic row approximation.\cite{papertheory} Therefore all $\mathbf{q}$ vectors lie within one plane in the reciprocal space and a vector product of any two of them is normal to this plane - i.e.\ all $\mathbf{Q}$ vectors have the \emph{same direction}.


An important feature of the EMCD sum-rules is the fact that these allow in principle any detector position to measure magnetic moments. I.e.\ the experiment is by no means limited to placing the detector on the Thales circle above two strongest Bragg spots.\cite{nature} This observation is particularly important for the new experimental geometries\cite{lacbed,lacdif} and for the optimization of the signal to noise ratio.\cite{snr}

\begin{figure}
  \includegraphics[width=8.5cm]{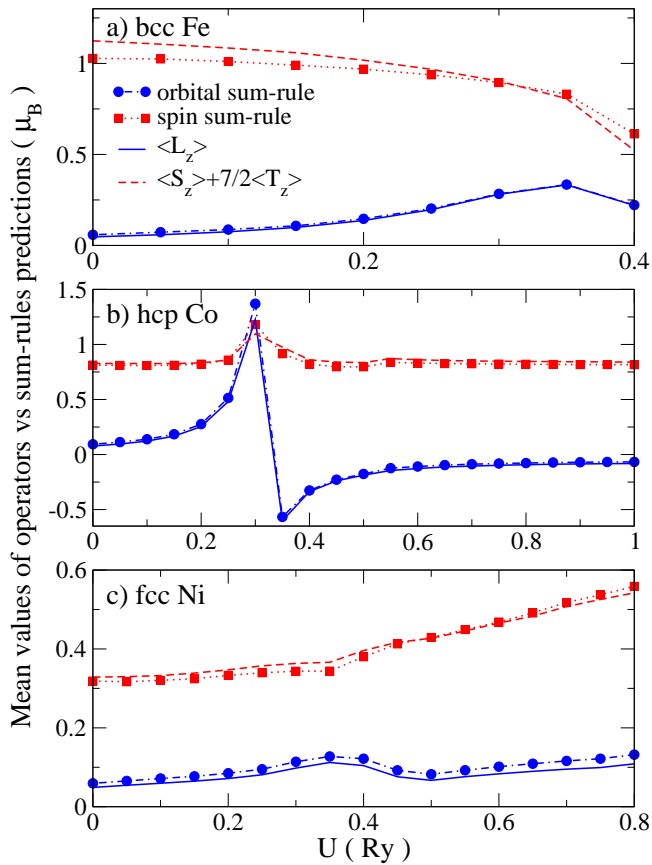}
  \caption{(color online) Comparison of {\it ab initio} orbital (blue solid line), spin and magnetic dipole moments (red dashed line) to the simulated sum-rules prediction (symbols) as a function of effective $U$ parameter (see text).\label{fig:sr}}
\end{figure}

The accuracy of the derived sum-rules naturally needs to be evaluated. We have investigated this by comparing the spin and orbital moments obtained from the EMCD sum-rules for a complete simulation of the dichroic signal\cite{papertheory} of bcc Fe, hcp Co, and fcc Ni, to exact spin and orbital moments. The latter values are calculated from the quantum mechanical expectation value of the spin and angular momentum operator, using wavefunctions from first principles theory. In the past such calculations have been shown to agree very well with measured moments of e.g.\ Fe, Co, and Ni.\cite{soderlind} 
In our calculations\cite{calcs} we have employed the strength of the electron-electron interaction (represented by the so-called Hubbard $U$ parameter) as a means to vary the size of the spin and orbital moments. Figure \ref{fig:sr} demonstrates an excellent agreement between the moments extracted from the sum-rules and the calculated operator mean values. 
It should be noted that in general the value of $\langle T_z \rangle$ is low, but for hcp Co in the range $0.2 \le U \le 0.4$~Ry it attains a rather large value. This is caused by changes in the electronic structure and will be analyzed in detail elsewhere.

Finally we note that an actual application of the EMCD sum-rules to extract absolute number of spin and orbital moments is more difficult than in XMCD experiments. Due to the effects of the dynamical electron diffraction the sum-rule analysis relies on knowledge of the sample thickness dependent function, which enters as a different prefactor within imaginary and real parts of the MDFF's, respectively. This can be calculated using the dynamical diffraction theory.\cite{papertheory} 

On the other hand, the \textit{ratio} of spin and orbital moments can be extracted in a completely analogous way as in the XMCD, because both spin and orbital momentum sum-rules depend on the {\it same prefactor} (up to a factor $3/2$). Similarly if the `extra terms' in $N$ and spin-orbit sum-rules are negligible, the ratio of $\langle \mathbf{L}\cdot\mathbf{S} \rangle$ and $N_e$ can be obtained.

In conclusion, we have derived a complete set of sum-rules for the general ELNES experiment. Our theory provides a foundation for the use of the EMCD technique as a quantitative tool in magnetic studies. Such quantitative EMCD analysis will open up new avenues for research in the field of nano-magnetism, owing to an excellent spatial resolution in combination with atom-specificity, primarily for obtaining information about spin and orbital moments.

This work has been supported by the European Commission, contract nr. 508971 (CHIRALTEM). We acknowledge financial support of STINT, the Swedish Research Council and the Foundation for Strategic Research. 


\begin{thebibliography}{99}
\bibitem{sysrow} Within a systematic row approximation also the original experimental geometry used in Ref.~\onlinecite{nature} was very close to such symmetrical conditions.
\bibitem{calcs} Electronic structure calculations were done using the FP-LAPW method with 40000 or 20000 $\mathbf{k}$-points in the first Brillouin zone for Fe, Ni or Co, respectively. Atomic sphere radii were set to 2.2 Bohr radii for Fe and Ni and 2.3 Bohr radii for Co, respectively. Approximately 100 basis functions per atom were used.
\end{thebibliography}
\end{document}